%% file: main.tex
\documentclass[sigconf]{acmart}

\AtBeginDocument{%
  }

\copyrightyear{2026}
\acmYear{2026}
\setcopyright{cc}
\setcctype{by}
\acmConference[CHI '26]{Proceedings of the 2026 CHI Conference on Human Factors in Computing Systems}{April 13--17, 2026}{Barcelona, Spain}
\acmBooktitle{Proceedings of the 2026 CHI Conference on Human Factors in Computing Systems (CHI '26), April 13--17, 2026, Barcelona, Spain}
\acmPrice{}
\acmDOI{10.1145/3772318.3791586}
\acmISBN{979-8-4007-2278-3/2026/04}

\begin{document}

\title{Transient Non-Use: How People in Migration Experience Digital Disconnection}

\author{Jonathan Leuenberger}
\email{jonleuen@nmsu.edu}
\affiliation{%
    \institution{New Mexico State University}
    \city{Las Cruces}
    \city{New Mexico}
    \country{USA}
    }

\author{Anamika Rajendran}
\email{arajen97@nmsu.edu}
\affiliation{%
    \institution{New Mexico State University}
    \city{Las Cruces}
    \city{New Mexico}
    \country{USA}
    }

\author{Augusto Penzo Jara}
\email{augustoj@nmsu.edu}
\affiliation{%
    \institution{New Mexico State University}
    \city{Las Cruces}
    \city{New Mexico}
    \country{USA}
    }

\author{Tajwar-Ul Hoque}
\email{tajwarh@nmsu.edu}
\affiliation{%
    \institution{New Mexico State University}
    \city{Las Cruces}
    \city{New Mexico}
    \country{USA}
    }

\author{Shiva Darian}
\email{shiva@nmsu.edu}
\affiliation{%
    \institution{New Mexico State University}
    \city{Las Cruces}
    \city{New Mexico}
    \country{USA}
    }


\renewcommand{\shortauthors}{Leuenberger et al.}
\begin{abstract}
People experiencing migration endure many transitions across borders, technologies, and social systems. While HCI research often emphasizes this community's adoption of technology, less attention has been paid to practices of technological non-use. This paper investigates how information and communication technologies (ICTs) are intentionally and unintentionally avoided, withheld, or not used during migration. Drawing on interviews with 32 people experiencing migration in the border city of El Paso, Texas, USA between February and May 2025, we identify a range of non-use experiences, including device, informational, and protective non-use. We extend the concept of non-use by situating it within the three phases of transitions: understanding, negotiating, and resolving. We show how ICT non-use shifts with time, risk, and institutional demands. Our analysis demonstrates that non-use functions both as a protective strategy and as a response to systemic exclusion, and concludes with design principles that anticipate non-use as both intentional and unintentional design conditions rather than as punitive failure. 
\end{abstract}



\maketitle
\input{sections/1_intro}
\input{sections/2_lit}
\input{sections/3_methods}

\input{sections/4_nonUse}
\input{sections/5_discussion}

\input{sections/6_conclude}
\bibliographystyle{ACM-Reference-Format}
\bibliography{ref}
\appendix
\input{sections/7_appendix}
\end{document}

%% file: sections/1_intro.tex
\section{Introduction}

Along uncertain migration routes, at border crossings, in detention centers, and in shelters along the way, phones are often described as a 'lifeline'~\cite{maitland2018digital}. A single device can connect people on the move to family~\cite{gillespie2018syrian,newell2016information}, aid~\cite{Talhouk2016Syrian, Talhouk2017implication,latonero2018digital}, navigation~\cite{merisalo2020digital}, and critical information~\cite{newell2016information}. Yet just as often, ICTs are absent, refused, surveilled, or dangerous to use. We contribute to a deeper understanding of \textit{non-use}, discussing how sociopolitical systems, surveillance structures, power dynamics, physical terrain, and other risks shape technological interactions. For HCI, \textit{non-use} offers an insightful perspective on how technologies embedded in systems of power are intertwined with the experiences of people undergoing migration.

Human migration is not a marginal or exceptional phenomenon but a persistent and global feature of contemporary life~\cite{abel2014quantifying}.
People move across borders for many reasons, including conflict, climate change, economic precarity, family reunification, and linguistic or geographic proximity~\cite{chen2009perceived}. Some journeys are planned and documented; others are improvised, indirect, or shaped by rapidly changing policies and border regimes. Migration routes frequently pass through transit cities and temporary destinations~\cite{castagnone2011transit}, where people may stay for days, months, or longer while waiting, working, or reassessing their options. These extended periods of uncertainty fundamentally shape how technologies are used, avoided, shared, abandoned, or withheld.

In migration contexts, non-use of ICTs is not simply a matter of preference or lack of access. It is often a rational response to risk. Phones can expose location, social ties, or past movements; they may be searched, confiscated, or used as evidence in encounters with authorities~\cite{newell2016information}. Connectivity itself can be uneven, expensive, or dependent on informal infrastructures such as shared devices, public Wi-Fi, or local aid organizations. As a result, people on the move frequently decide when, how, and whether to use digital technologies at all \cite{molnar2024walls, steinbrink2021europeprivacy}. Examining non-use allows us to understand how people actively navigate uncertainty and power through acts of refusal, rather than framing people as passive recipients of technology or policy.

HCI scholarship has long engaged with experiences of people on the move, emphasizing how ICTs facilitate connection~\cite{vernon2016connecting}, information seeking~\cite{newell2016information}, and adaptation~\cite{Xu2015Promoting} in displacement. Studies have documented the role of smartphones in sustaining international ties~\cite{latonero2018digital}, enabling navigation~\cite{merisalo2020digital}, and supporting integration~\cite{semaan2016transition}. However, this focus on access and adoption risks overlooking the conditions under which ICTs are not used. There is some existing work on non-use studies, for example, Satchell and Dourish's foundational typology of non-use highlights how disengagement can range from deliberate (through active resistance, disinterest, or lagging adoption) or involuntary (through disenfranchisement, disenchantment, or displacement)~\cite{satchell2009beyond}. For people on the move, these dynamics take on heightened stakes: intentional acts of disconnection such as turning off a phone may protect them from potential violence, while systemic exclusion may cut off access to digital infrastructures altogether. Non-use, therefore, is central to how people experiencing migration engage with technology. 


We use the terms \textit{people on the move,} \textit{people crossing borders,} and \textit{people experiencing migration}, following Molnar's call to expand the use of people-first terminology when discussing human migration and lived experience~\cite{molnar2024walls}. These terms refer to the breadth of people who may or may not fall under rigid legal and policy categories such as `refugee,' `asylum seeker,' and `migrant.'

People on the move are individuals who have left (or been forced to leave) their country of origin and are navigating uncertain, often prolonged journeys across multiple national contexts. These journeys are frequently shaped by violence, exploitation, legal precarity, and shifting border regimes, regardless of the eventual destination \cite{molnar2024walls}. Across regions, people experiencing migration often encounter structural marginalization, including restricted access to housing, healthcare, and legal protection, or heightened exposure to surveillance and enforcement at borders and within host countries \cite{atak2024vulnerability}.

People on the move experience drastic changes in their socio-political identity~\cite{cormoș2022processes}. They are navigating through \textit{transitions}, movements between life stages that involve changes to identity and behavior~\cite{ruthven2022information, willson2019transitions}. Previous work studying transitions have discussed contexts such as parenting~\cite{ruthven2018relationships}, education~\cite{ecclestone2009transitions}, and health~\cite{meleis2000experiencing}, using transitions as a method to explain these changing life stages. We utilize Ruthven's transitions theory which distinguishes the transition based on the psychological stages of the person experiencing it and the subsequent information seeking strategies in each stage (understanding, negotiating, and resolving). Understanding how and why ICT's are \textit{not} used during the migration journey reveals how sociotechnical systems affect and shape psychological processes and limits of mobility. 

Our study draws on 30 semi-structured interviews with 32 people on the move who have experienced migration through the US-MX border. Interviews were conducted in El Paso, Texas, USA between February and May 2025, during a period of intensified changes in US immigration policy. In this paper, we focus on how people on the move navigate ICT use and non-use within these transitions and how circumstantial, governmental, and technological infrastructures configure conditions of possibility. In doing so, we extend HCI's engagement with non-use by situating it in the context of forced migration, where non-use (especially deliberate non-use) is not a deficit, but a meaningful way of engaging with institutions and society. We ask the following research questions:
\begin{enumerate}
    \item How do people on the move experience and make sense of moments of unintentional non-use of ICTs?
    \item How and why do people on the move sometimes intentionally avoid or refuse the use of ICTs or sociotechnical systems?
    \item What practices, strategies, or workarounds do people on the move develop when ICTs are unavailable, inaccessible, or unsafe? 
\end{enumerate}

In what follows, we review HCI literature related to ICT use and non-use, particularly related to vulnerable communities, before outlining our methodological perspectives. We then present our findings about how people in migration experience ICT non-use during their journeys to the US. Finally, we discuss non-use within the context of transitions and offer design recommendations that account for ICT non-use at different stages of the journey.

%% file: sections/2_lit.tex
\section{Literature Review}

ICTs are central to daily life, providing access to information, connection, and problem-solving. For people in migration, they are vital for navigating uncertainty, securing resources, and sustaining ties across borders. Research approaches these issues through two lenses: \textit{use}, which examines adoption and adaptation, and \textit{non-use}, which highlights disengagement, resistance, or barriers to access.

\subsection{ICT Use}
ICT \textit{use} refers to how individuals engage with digital technologies to access information, maintain social connections, and navigate daily life or transitional circumstances. For people in migration, factors such as facing the uncertainty of the journey, securing resources, and sustaining social ties all shape the way ICTs are used. 

Access to mobile devices are vital for people in migration worldwide~\cite{brunwasser201521st}, particularly as governments rely on app-based technology to facilitate border crossing procedures (e.g., the Customs and Border Patrol (CBP) App in the US~\cite{Montoya-Galvez2024}, Europe’s MigApp~\cite{IOM_MigApp}). Access to both cellular data and mobile devices is critical to the well-being, personal enjoyment, and mental health of people experiencing migration~\cite{vernon2016connecting}. As governmental systems and societies increasingly rely on technology, human migration has become more dependent on digital technologies, prompting a new understanding of the journey as a “digital passage,” where navigating social connections and infrastructures parallels the physical journey~\cite{latonero2018digital}.

Understanding this digital passage, including its many stakeholders (people experiencing migration, aid workers, lawyers, cartels, law and immigration enforcement) offers a socio-digital lens on digital divides, liminality, and the transitions undergone by people experiencing migration. Chang and Gomes conceptualize this passage with their concept of the ``digital bundle,'' the collection of websites, apps, and online information sources that people in migration rely on~\cite{chang2019digital}. This digital passage can be understood as the movement from one such bundle of information sources to another, illustrating how migration is a journey both mediated by technology use and a journey toward new technological ecologies.  

ICTs play a central role during transitional periods, helping individuals adapt to uncertainty, maintain social connections, and access critical resources. In the context of migration, studies show that people on the move rely on digital tools to access news and information~\cite{gillespie2018syrian, Wall2017syrian}, to sustain themselves financially~\cite{betts2016Refugee}, to access health services~\cite{Talhouk2016Syrian, Talhouk2017implication}, and to maintain mental well-being~\cite{Mikal2015Refugees}. Design-oriented research further highlights how information infrastructures and digital media can help people experiencing migration maintain identity~\cite{Fisher2017Za'atari}, imagine futures~\cite{Fisher2016Future's}, and navigate social and spatial transitions~\cite{fisher2019osool, Xu2015Promoting}. Together, these studies illustrate how the use of ICT supports adaptation, resilience, and coping during periods of change.

\subsection{ICT Non-use}
While ICTs offer critical support and adaptability during transitions, access to and engagement with these technologies is not uniform; barriers, constraints, and deliberate choices also shape how, when, and whether individuals use ICTs, highlighting the importance of examining ICT non-use. Previous research on non-use has examined who does not use technology, why, and with what consequences~\cite{selwyn2004reconsidering,warschauer2004technology}.

Satchell and Dourish~\cite{satchell2009beyond} identified six forms of non-use (displacement, disenchantment, disenfranchisement, lagging adoption, disinterest, and active resistance), which can be broadly understood along a spectrum from unintentional to intentional non-use, with some behaviors lying in between. Unintentional non-use occurs in situations where a person experiencing migration desires to engage with technology but is restricted by external circumstances; this exemplifies disenfranchisement and displacement, but can also include disenchantment. Intentional non-use reflects deliberate decisions to disengage from technology and encompass active resistance, disinterest, and lagging adoption. 

\subsubsection{Unintentional Non-Use}
Unintentional non-use arises from structural and systemic barriers that prevent equitable access to technology, or ``digital divides''~\cite{van2006digital}. These divides are shaped by socio-demographic factors such as gender, age, education, economic status, and geography; they limit access and benefits from ICTs~\cite{haile2021liberalities}. For instance, women and under-educated people experiencing migration are disproportionately less likely to own or have regular access to mobile phones, limiting their ability to connect, access information, and maintain social ties~\cite{brunwasser201521st,kabbar2006factors,van2014digital}. Discrete studies on undocumented people at the US–Mexico border found that storing friends’ and relatives’ contact information increased one's vulnerability to extortion and exploitation by drug traffickers, human smugglers, thieves, and corrupt authorities~\cite{newell2016information}. 

Digital divides have been categorized into three levels: first-level divides involve physical access to devices and connectivity~\cite{van2019first}; second-level divides relate to social and material inequities limiting access; and third-level divides involve digital literacy, shaping the ability to effectively use available technology~\cite{van2015third,chinn2004determinants}. People experiencing migration are impacted across all three levels, resulting in limited information access and constrained use of ICTs~\cite{dryden2011refugee,alam2015digital}. These situations reflect disenfranchisement, disenchantment, or displacement~\cite{satchell2009beyond}, where non-use is imposed by circumstances rather than choice.

\subsubsection{Intentional Non-Use}
When people in migration engage in intentional non-use, they often deliberately limit ICT use due to perceived risks. People experiencing migration often rely on mobile phones for various purposes, but they also worry that these devices could increase their vulnerability to extortion and expose them electronically to what they perceive as highly advanced governmental surveillance systems~\cite{newell2016information}. Concerns about extortion, surveillance, financial cost, or exposure to authorities lead people experiencing migration to selectively disengage from technology, reflecting active resistance, disinterest, or lagging adoption as described by Satchell and Dourish~\cite{satchell2009beyond}.

\subsection{Transition Stages}
As people on the move progress through their migration transitions, they experience distinct informational needs and behaviors. Using Ruthven’s three-phase model of transition---understanding, negotiating, and resolving---we examine how ICT (non-)use shape movement, decision-making, and survival during migration. These phases occur repeatedly as individuals confront new environments, threats, or legal uncertainties. 

The earliest transition aligns with Ruthven’s notion of \textit{understanding}, when people experiencing migration begin to recognize that they must leave their current living situations~\cite{ruthven2022information}. While coming to this recognition, understanding serves as a cognitive and emotional orientation in which people construct an initial model of what migration may entail before departing~\cite{ruthven2022information}. 
During this period, individuals search for information by contacting those who have previously migrated, evaluating possible routes, or identifying perceived risks and opportunities~\cite{du2023understanding,bahl2025whole,nikkhah2020coming}. This phase is characterized by uncertainty, dissonance, and the sense that current conditions are untenable. 

Once individuals commit to movement, they transition into \textit{negotiating}, where they manage their situation to minimize disruption and effort. This is enacted through travel, logistical coordination, creating connections with individuals of shared backgrounds and interaction with various legal and informal systems~\cite{jung2025transitions}.

After negotiating, people experiencing migration begin \textit{resolving}, where they attempt to stabilize their situation both within new institutional configurations and their search for meaning within a new country.  Upon arrival at the U.S.–Mexico border, people experiencing migration entered periods of adapting, when they attempted to stabilize their situation within new institutional configurations. This takes the form of legal encounters, ongoing adjustments to information systems, appointments, expectations, and documentation requirements~\cite{chang2019digital}.

While ICT use by people in migration is well-studied in European and Asian contexts, there is limited research on technology use during migration across the US-MX border, which is the focus of this research. Research on technology non-use in the high-risk, in-transit experiences of people undergoing migration into the US through Mexico, is minimal, with the notable exception of work by Newell in 2016~\cite{newell2016information}. Our study addresses this gap by analyzing  ICT experiences across transition phases, highlighting how people's actions reflect patterned, meaningful responses to power and precarity. By connecting Ruthven's information behavior theory of transitions with Satchell and Dourish's topology of non-use, we show how these responses emerge within conditions of uncertainty. We further discuss how HCI can design systems that anticipate disconnection, support reconnection following periods of non-use, and treat refusal itself as a considered design decision.

%% file: sections/3_methods.tex
\section{Research Context}
This work was conducted in El Paso, Texas, USA, a city that shares a border with Ciudad Juárez, Chihuahua, MX. The cities are a central part of the Paso del Norte region, serving as a juncture in the northern passage for people on the move. Both cities have share a longstanding network of aid organizations accommodating those passing through. 

Data were collected between February - May 2025, immediately following the second inauguration of US President Donald Trump. During this period, significant policy and operational changes led many El Paso aid services to close, redirect resources to people who had entered the US prior to the change in administration, or to accommodate those who were newly released by US Immigration and Customs Enforcement (ICE) after residing in inland cities across the US for many years. 

\input{figures/participantDemographics}
\section{Methods}
This interview protocol is informed by nearly three years of community-based engagement by the final author, who has a background in refugee resettlement, and is a child of immigrants and speaks English, Spanish, and Farsi. This work foregrounds methods of \textit{accompaniment}~\cite{yarris2024accompaniment} by centering participation in the community through volunteer work, attendance at community events, and interviews with employees, volunteers, and activists across the local network of immigration-oriented organizations. 

This study is part of a larger, exploratory interview protocol about individuals’ experiences in migration that was co-developed with local community activists and distributed across the network for feedback from community leaders focused on migration. The overall protocol questions focused on how individuals prepared for migration, experiences with officials (including governmental, organizational, and informal entities), and how they navigated online information and communication. This particular work is primarily focused on references to ICT non-use and coping strategies disclosed by participants throughout the duration of the interview. The remaining portion of the interviews provides additional context and insights. The protocol themes can be found in Appendix ~\ref{appendix:protocol}

\subsection{Participant Sampling}
We recruited participants through flyers posted at communal locations in El Paso (e.g., shelters, parks, cafes, bus and train stations, near ports of entry). Flyers were also circulated via private group chats used by local nonprofit and government workers coordinating services for people in migration. Flyers recruited people who self-identified as being in migration. 
Community partners supported safe recruitment by allowing community-based interviewers and one member of the research team to make announcements at local immigrant-serving nonprofits. Community-based interviewers with proper research credentials confirmed participant eligibility while coordinating interview times, briefly asked participants to confirm their age and country of birth, and to provide a short reason for what motivated their migration. This question was meant to filter out anyone who identified as being bi-national, living in both US and MX cities in the El Paso - Cd. Juárez borderplex. 

\subsection{Participants}
\label{participant_demographics}
Across all recruitment channels, 44 individuals responded to flyers and outreach efforts. Of these, 36 individuals scheduled and participated in an interview. Five individuals were screened out when they were identified as being bi-national or who declined to participate or coordinate/schedule an interview after learning more about the study procedures. While recency of migration experience was also considered, those who were managing Alternative To Detention (ATD) surveillance devices (e.g., ankle monitors, smartwatches, mobile phones) assigned to them by the US Department of Homeland Security (DHS\footnote{DHS is comprised of many entities, including Customs and Border Protection (CBP), Immigration and Customs Enforcement (ICE), and the Transportation Security Administration (TSA). In this work, we refer broadly to 'DHS' so as not erroneously attribute participant experiences to a specific entity.}) were included. Two interviews were stopped early due to discomfort of research team members.  

Among the interviews conducted with 33 individuals, one interview is excluded from this dataset due to the participant having a highly unique experience compared to the rest of the participants, crossing as a child and living under Deferred Action for Childhood Arrivals (DACA). Thus, this study focuses on 30 interviews with 32 individuals experiencing migration. These in person interviews lasted between 30-75 minutes (mean 54 minutes) in El Paso. There were two joint interviews with individuals who had crossed borders together. 

Data saturation was reached as each new interview was no longer adding unique insight into ICT non-use experiences. The research team stopped conducting in-person interviews after they shared several rounds of announcements across the local aid-organization network. At that point, they were confident that all likely participants had been informed and that those interested had adequate opportunities to take part.

The vast majority of participants were originally from Venezuela; fewer participants were originally from Mexico and Ecuador, and even fewer participants were from other nations\footnote{To avoid compromising participant anonymity, we do not provide cross-sectional information.}. At the time of interview collection, there was a limited number of people in migration physically located in El Paso. We present Table~\ref{tab:participant_overview} as a disconnected summary of participant demographics and status categories, while preserving participant privacy. 

All participants in this study crossed the border prior to the 2025 US Presidential inauguration. Most participants had recently completed their transition journeys, and had been in the US for 0-3 years. Although 3 had been in the US for longer than 10 years. Numerous participants had lived in other cities within the US, coming to El Paso for work, the aid network, or to self-deport or otherwise repatriate.  

\subsection{Interview Procedure}
Interviews were conducted by the final author and community partners, all of whom completed institutional ethics certification. Interviews were held in quiet public locations (e.g., parks, cafes, shelter lobbies) and conducted in participants' preferred language. The interviewers were all active in the local network of immigrant-serving community based organizations. All participants received \$20 cash compensation after consenting to participate and be audio-recorded for the study prior to beginning the interview.  We also documented surveillance devices (e.g., noting (occasionally photographing) ATD technologies and contracted providers). This study was approved by the university's institutional ethics board (IRB).

Because most of the interviewers a) did not have formal training in qualitative interview methods outside of the ethics certification required by the university, b) occupied different positions throughout the community (e.g., volunteer, community organizer, researcher who engages in volunteer work), and c) had different positionalities (including lived experience with migration, as someone from border region, as an immigrant, as a child of immigrants), reflexive discussions occurred before and after interviews. The community-based researcher conducted 3-5 interviews alongside each community interviewer before they conducted interviews alone. The community-based researcher would give feedback about interviewing techniques after listening to recordings and joined for interviews on an occasional basis following initial training.  

Due to the sensitive nature of the research, interviewers explicitly emphasized participant control (e.g., skipping questions, pausing the interview, or declining sensitive topics) numerous times to reduce power differentials. Interviewers used trauma-informed practices~\cite{isobel2021trauma} to navigate sensitive topics (e.g., experiences of violence, documentation status, trauma). Interviewers avoided asking about traumatic details during periods of disconnection. If a participant showed signs of distress, interviewers paused the session, allowing them to rest, and offered to end the interview. 

\subsection{Transcription Verification and Data Security}
Audio recordings, notes, and photographs were stored on a secure private server. The second author used a self-hostable AI audio transcription and translation tool to ensure that participant data was not shared with third parties. A locally hosted instance of the WhisperX model \cite{whisperx} was used primarily for transcription with word-level alignment and speaker diarization. A custom script extended the workflow to also perform translation into English and generate speaker-differentiated transcripts. Background noise reduction was applied as needed using local audio software, DaVinci Resolve \cite{davinci}, before processing. The resulting transcripts were saved as text files with speaker labels coded for clarity and were uploaded to the same secure local server used by the research team, allowing the third author to review and verify the content while maintaining strict data privacy. To prioritize participant anonymity, any references to immigration status, or identifying details (e.g., locations, names) were deidentified during transcription verification, swapped with relational descriptions.

\subsection{Analysis}
We analyzed interviews using inductive open coding~\cite{strauss1994grounded} and thematic analysis~\cite{patton2015qualitative}. Initial codes were developed iteratively from analysis for emergent themes from the transcripts. The first two and final two authors coded the data individually, while the third author, who oversaw translation verification, wrote reflexive memos following each transcript review. The team met regularly to discuss interviews, review codes, and compare annotations. When discrepancies arose, they were resolved through discussion and memoing. 

Our analysis focused on how participants experienced and navigated ICTs during their migration journeys, particularly in contexts of limited access, surveillance, or perceived risk. Through this process, we identified recurring patterns of ICT non-use. These patterns reflect a range of intentional and unintentional strategies participants employed to manage safety, privacy, and access, highlighting how non-use is an active and contextually shaped aspect of digital engagement rather than a simple absence of technology.

%% file: figures/participantDemographics.tex
\begin{table*}[t]
\centering
\caption{Participant demographics: Overview of gender, age, country of origin, and participant type.}
\begin{tabular}{p{4cm}p{8cm}r}
\toprule
\textbf{Category} & \textbf{Subcategory} & \textbf{Count} \\
\midrule
Gender & Female & 14 \\
       & Male & 18 \\
\midrule
Age (years) & 19--29 & 11 \\
            & 30--39 & 15 \\
            & 40+ & 6 \\
\midrule
Country of Origin 
& Venezuela, Mexico, Ecuador, Guatemala, Colombia, Nigeria, Kenya 
& 32 \\
\midrule
Participant Type & Single Adult & 16 \\
                 & Family Unit & 16 \\
\bottomrule
\end{tabular}
\label{tab:participant_overview}
\Description{This table shows participant demographics. The left table displays gender (14 females, 18 males) and age groups (11 aged 19--29, 15 aged 30--39, 6 aged 40+). The right table lists countries of origin (Venezuela, Mexico, Ecuador, Guatemala, Colombia, Nigeria, Kenya), and participant type (16 single adults, 16 family units).}
\end{table*}

%% file: sections/4_nonUse.tex
\section{Findings}
Our analysis reveals how people undergoing migration experience three main types of non-use.
\begin{enumerate}
    \item Device Non-use
    \item Informational Non-use
    \item Protective Non-use
\end{enumerate} In what follows, we detail these forms of non-use and share how people experiencing migration persevere through periods of non-use. 

Participants described mobile phones and internet access as digital lifelines that facilitate communication with family, access to information, and navigation through journeys and immigration procedures. Yet, prior research shows that device possession can also introduce vulnerability~\cite{newell2016information}. As governments increasingly digitize immigration processes, mobile devices become both more valuable and more risky: people experiencing migration often need a mobile phone to navigate entry into the US and maintain contact with family, yet fear that carrying a device may expose them to surveillance, targeting, or exploitation. 

\subsection{Device Non-Use (Loss of Device)}
Device non-use includes moments where people experiencing migration lose access to their phones due to damage, theft, confiscation, or disconnection. Phones are critical to nearly every stage of the journey. When a device is lost, so too is access to information, networks, and occasionally, even proof of identity. For people experiencing migration, periods of non-use often arise from unintentional (sometimes anticipated) disruptions. Phones are constantly at risk of being lost (P19), damaged (P11), or stolen (P20) during the journey. Even when devices remain intact, participants still face disconnection through irregular cellular coverage (P18), border crossings (P5), and deliberate restrictions imposed by authorities, cartels (P6, P18), or governments that deliberately remove or limit phone access. In each case, the result is extended and often difficult periods of forced disconnection from both family and financial support as well as bureaucratic processes. 

\subsubsection{Loss, Damage, Theft, and Confiscation}
Phones were frequently damaged en route. Some participants recounted devices breaking during the journey through the jungle, while others described water damage or accidents during river crossings. P4, for example, recalled how their spouse's phone ``\textit{didn't serve (us) that much...because he got it wet when he threw himself into the river.''} Others described phones being stolen outright, and for many, theft felt normalized. ``\textit{Right now I don’t have a phone because they stole it here where I’m staying...it’s been about a month since they stole it...they steal everything from me there}'' (P20).
In some cases (such as P21 and P29), soon after someone experienced phone theft, their families were contacted for false ransom payments, despite no one actually being held captive. In such cases, the absence of a phone not only resulted in disconnection, it also enables new forms of coercion. Without a device to verify their safety, families became vulnerable to extortion schemes. 

Circumstantial non-use also included systematic restrictions imposed by more powerful stakeholders, notably, foreign law or immigration enforcement and smugglers or cartel members. Participants described how such stakeholders weaponized the removal and restriction of ICTs to limit internet access (P1, P18), translation services (P3, P4, P5), their ability to connect with family or legal counsel (P5, P18, P29, P32), and their capacity to pursue asylum (P10, P11). Throughout their journey, P12, for example, experienced restrictions imposed by private groups, NGOs, and governments on internet access. In P12's experience, the barrier to access was financial cost.
\begin{quote}
    \textit{Interviewer}: And how did you communicate with your [spouse]?
    \\
    \textit{P12}: by WhatsApp, but when we arrived at the first camp, there was Wi-Fi, but we always had to pay to use the Wi-Fi, 
\end{quote}

Many participants additionally shared their firsthand accounts of ICTs being used as tools of coercion during kidnapping. When kidnapped, many participants recalled being forced to disclose contact information for ransom, to allow kidnappers to search their phones, or to surrender their devices entirely. As P5, explained, ``\textit{I was close to Piedras Negras but I had not yet reached the border. Which was when they kidnapped us. There I lost the phone, I lost everything}.'' For P5, captivity turned into enforced non-use, where losing the phone meant `losing everything.' This underscores how phone seizure was not just material theft, but an existential interruption, temporarily stopping their capacity to navigate, seek help, or maintain connection. Non-use is not just a technical absence but an imposed condition, enforced through recurring exploitation that stripped participants of their abilities to maintain digital ties.

In the US, these restrictions were most visible while in with DHS officials. Most of our participants reported being barred from using phones while waiting to be processed by DHS entities, even when contact with family or legal support would have resolved critical circumstances. P3, who was asked to prove their paternity without internet access shared, 
\begin{quote}
    \textit{P3}: There are like 30 to 40 people in a single room they have given to everyone. 
    \\
    \textit{Interviewer}: And is there internet? 
    \\
    \textit{P3}: No.
    \\
    \textit{Interviewer}: At this moment are there people using their cell phones? 
    \\
    \textit{P3}: No, there is no one. Absolutely no one goes with cell phones.
\end{quote}
Here, disconnection was institutionalized, enacted as a hostile policy that silenced detainees, preventing them from seeking legal help or reassuring their families. Taken together, these accounts show how non-use through loss, theft, and confiscation is not only about device fragility but about power. Phones were lost to rivers, jungles, and thieves, but also strategically withheld by cartels, officials, and detention systems. Each episode of non-use underscores the precariousness of digital dependence under displacement, revealing how disconnection itself becomes a mechanism of control and coercion.

\subsubsection{Infrastructural Non-Use}
Participants described connectivity challenges emerging from infrastructural gaps. Periods of disconnection ranged, depending on the size and athleticism of the groups, as well as luck about the conditions (e.g., weather, natural disasters) that people encountered. Groups with families, elderly or pregnant travelers, moved more slowly, and, accordingly, endured longer periods of disconnection. Leaving settled areas often meant losing all access to internet and cellular networks. P13 recalls their experience in the jungle: \textit{``After we left the camp we had no more communication with anyone. No, after the camp it was just mountains''} (P13). These expected disconnections sometimes shaped strategies of non-use. Some participants chose not to buy SIM cards in countries they would only be in briefly, or where they anticipated that service would be limited or unavailable. ``\textit{Sometimes there was no signal, because like I said it was different countries, so we had to have different SIM cards for each country. So no, we always relied on Wi-Fi and that’s it}'' (P30). Such anticipatory adaptations show how people experiencing migration incorporated circumstantial non-use into their strategies, budgeting around predictable disconnection even as it heightened risk.

Taken together, these accounts highlight how non-use was often circumstantial and beyond individual control. Device loss, damage, theft, confiscation, as well as infrastructural disconnection, created conditions in which people experiencing migration were simultaneously dependent on, and excluded from, the sociotechnical systems meant to support their passage. 

\subsection{Informational Non-Use}
Informational non-use emerges when people experiencing migration cannot, or choose not to, engage with informational resources or infrastructures of record-keeping, bureaucracy, and surveillance.

\subsubsection{Unintentional: Misunderstanding and Misinformation}
Informational non-use in this study often stemmed from confusion and misinformation across multiple powerful infrastructures. On one side, guides, smugglers, and cartels controlled access to information along the journey, often withholding details about routes, risks, and timing in ways that left many of our participants unprepared for border crossings, government checkpoints, and geographic dangers. On the other side,  participants who entered US government systems encountered complex and fragmented bureaucratic processes, where missing a meeting, misinterpreting a document, or misunderstanding a requirement could result in penalties, detention, or deportation. Together, these conditions undermined participants' ability to engage with ICT systems as intended, producing informational gaps that were rarely within their control. 

Many participants described feeling uncertain about what was happening during the journey, particularly where contracted guides, smugglers, and cartels controlled information about routes. Some reported not knowing they would need to climb the US-MX border wall to enter the US (P1), how long they would be held in stash houses throughout the journey before continuing (P6), or what sort of conditions they would have to endure along the journey (P12). These moments of withheld information forced many participants into states of unintentional non-use: unable to prepare, they could neither anticipate risks nor plan communication with family. P1, who used a guide found by their cousin, explains their experience learning that they would be expected to climb the US-MX border wall: ``\textit{Well, on the first try I didn't know the height...I didn't even know I was going to cross the wall. So I didn't even ask, because I thought I was going to walk}'' (P1). For many like P1, their vulnerable dependency on someone with a lot of power exacerbated informational asymmetries by preventing them from using ICTs to anticipate, coordinate, or seek support. 

Within US enforcement, participants were expected to track reporting requirements across multiple agencies, including with DHS entities or Intensive Supervision Appearance Program (ISAP) check-ins and court dates. Several participants described missing appointments due to misunderstanding times, locations, or modalities. P14 believed their remote meetings, called ‘virtual check-ins,’ could be conducted from any location. They explained: 
\begin{quote}
    \textit{P14}: It said it’s a virtual office visit. But I don’t know, because they told me before...they told me it was a call. And then they called and told me it was in person. So they had already visited my house...So I didn’t understand.
\end{quote}

Due to this miscommunication, P14 was penalized for not being inside their home during the scheduled visit time. Some participants (P12, P16) also knew they needed to change their address with DHS but did not realize they were expected to update their address with each agency separately (e.g., ISAP, Immigration and Customs Enforcement (ICE), and courts). Others misunderstood expectations regarding the location they were supposed to report to. P12 explains this with their description of how their spouse was penalized. ``\textit{Today, he’s arrested again because, according to them...he had a deportation order for not showing up to court}'' (P12). This account highlights how ICT systems designed for compliance convert misunderstandings into violations, producing non-use through bureaucratic opacity. Instead of ICTs enabling coordination, our participants  were effectively excluded from meaningful engagement.

Language barriers further exacerbated these dynamics. When entering the country, several participants described US officials communicating only in English, even during high-stakes interviews and document signing. P5 recounted: 
\begin{quote}
    \textit{P5}: I was interviewed by a computer. And the [virtual] ICE officer who was interviewing me spoke English. And I did not understand what they were saying. And I did not have an interpreter. And the person (officer) who was [physically] with me spoke with the [virtual] ICE officer. And I did not know what was going on. And I see it as something inappropriate because they could incriminate me of something. I do not know what happened. I think there should be an interpreter. I think so...He did not tell me anything.
\end{quote}
Here, non-use was not a matter of personal choice but of institutional design. ICT-mediated processes (computer interviews, document processing) functioned only in English, effectively disenfranchising non-English speakers. Similarly, P4 explained that everyone they were with in DHS processing were pressured to sign documents without translation:
\begin{quote}
    \textit{P4}: The whole room would say, `also what did they explain to you?' `No, they did not explain anything' `nothing', `what did they say to you?' `I didn't understand', `what did they ask you?' `Nothing, just \textit{¡Firme!} (Sign!)' One is signing. One does not know what is signed, it is bad. 
\end{quote}
These accounts demonstrate how ICT-mediated systems can structurally exclude users by enforcing mandatory ICT interactions (signing legal forms) without any attempt to mitigate language barriers, resulting in instances of coerced non-use. Many of our participants described engaging materially by signing forms, but doing so without informational access deemed their engagement as hollow. 

Alongside linguistic exclusions, participants also described federal stakeholders withholding information during detention and transfer. P5, for example, described being transferred between US detention facilities numerous times without being informed about where they were going. 
\begin{quote}
    \textit{P5}: And you do not know what is going to happen. So you ask, `where are you going to send me?' They [DHS] have processed us, they are going to deport us, they throw us on the border. But nobody says anything. Nobody says anything. I even spent months locked up where I did not know what was going to happen. My family did not know anything about me. I could never call my relatives.
\end{quote}
Both people experiencing migration and their families were often excluded from informational systems that would have allowed communication. Here non-use extends beyond individuals to affect entire social networks. P5 continued to describe that they were transferred to two different detention centers after entering the country in south eastern Texas; first to El Paso, TX and then to Connecticut. Each time, they were not told where they were going, why, for how long, or whether they were being deported. Several participants reported having their phones confiscated during detention and processing by US officials, sometimes without being returned upon release. 

Across our study, unintentional non-use emerged through misunderstanding, misinformation, language barriers, institutional opacity, enforced disconnection, and delay. These conditions shaped how participants could engage with ICT systems and had significant consequences on their ability to participate in legal, social, and informational processes.

\subsubsection{Intentional: Avoiding Visibility by Federal Infrastructures}
To avoid government visibility in the US, several of our participants crossed the US-MX border without being detected and deliberately avoided making contact with US federal officials. Some participants were motivated to cross this way due to running out of money and support in Mexico. Others, however, discussed how their decision to not use CBP One app to cross the border was based on not wanting to be in more government systems. Through either past harm or fear of future harm they decided that to best protect themselves they wouldn't use the formal legalized way of entry and instead chose to seek the protection that anonymity from the US government could provide. P20 explains how they knew about CBP One but was thankful for having not used it:
\begin{quote}
    \textit{Interviewer}: Did you try to use the CBP One app? 
    \\
    \textit{P20}: No, no. Thank God no, because honestly I wouldn’t like to. 
    \\
    \textit{Interviewer}: Why not? 
    \\
    \textit{P20}: Because it would put me in more systems.	
\end{quote}
In addition to the intentional avoidance of government systems, some participants described avoiding digital backups of identity documents. While digital copies could replace lost paperwork, many preferred not to create them, citing privacy concerns (P6), lack of trust (P20), or simply, because they had not considered it (P23). 

This refusal to engage with official systems underscores how informational non-use is about navigating asymmetrical power. Non-use here was a way to survive and avoid an every increasing technological dependence at the border. Understanding that people in migration might want to avoid this dependence through physical documentation and crossing undetected is crucial to our understanding of the border as a whole and to what technology should allow or support users to do. 

\subsection{Protective Non-Use}
For people experiencing migration in this study, devices were simultaneously indispensable and precarious: crucial for survival, navigating bureaucracy, and staying connected with family, yet also potentially exposing them to further harm and exploitation. To manage this tension, participants who did keep their devices often engaged in various forms of \textit{protective non-use}---intentional practices aimed at preserving privacy and safety. 

Protective non-use encompassed a variety of strategies, including logging out of shared devices, limiting online interactions, withholding information from family, deleting sensitive data before border crossings, and avoiding digital backups. Each practice highlights how everyday digital habits are recalibrated under conditions of precarity, where refusal is about minimizing risk rather than rejecting technology. The following subsections examine the strategies participants used to maintain control over their digital presence.

\subsubsection{Logging Out}
Borrowing and lending devices was common, and many participants described regularly logging into social media sites to avoid having to memorize phone numbers. Some participants described deliberately choosing platforms that allowed easy login and logout when borrowing phones, using account-based access as a way to maintain communication without long-term device ownership. For example, P4 explained why (Meta/Facebook) Messenger was preferred over WhatsApp in these situations, ``\textit{When I did not have WhatsApp we borrowed a phone and it was for Messenger that my mom is always connected because she knew that we did not have this phone}'' (P4). 
Elongated periods of disconnection, however, could mean permanent loss of access to accounts. P28 discussed how they resorted to using Facebook Messenger only if WhatsApp was not an option during circumstances such as losing a contact's phone number:   ``\textit{Every time I lost his number, I switched from WhatsApp and we’d communicate through Facebook}'' (P28).

\subsubsection{Limiting Online Interactions} 
Several participants described limiting their engagement with platforms to avoid stress, violence, or punitive repercussions. Some limited their exposure to violent videos and stressful news streams, while others avoided posting out of fear of legal processes. When asked about their social media consumption, one participant shared, ``I try not to pay much attention to it because I say to myself that I’ll go crazy if I do'' (P24). P6, instead describes not posting due to concerns over potentially punitive actions by federal entities:
\begin{quote}
    \textit{P6}: I try to be paranoid when I am going to comment something. I comment it without (sounding) against the President (Trump) or Elon Musk because it attacks...I always look for ways to form an opinion that is true to what I want to say but without getting into controversies because when I have my papers reviewed, [the agent] will say, `Oh yes, look at this tweet you said we were bad' [but] I never said that.
\end{quote}
For P6, caution became a survival tactic, showing how political fear shaped even mundane interactions. 

\subsubsection{Non-disclosure}
Non-disclosure emerged as a deliberate strategy used by participants to manage the complex reliance and expectations in familial relationships. Multiple participants explained that they hid their location or concealed hardships from their families to avoid causing their families stress, or to avoid disappointing expectations. 
\begin{quote}
    \textit{Interviewer}: When you talked to [family] what did you say to them?
    \\
    \textit{P21}: I'd tell them that I was doing fine. That I didn't sleep in the street, that I was at a luxury hotel, that it's easier now. 
    \\
    \textit{Interviewer}: Do they know where you are, where you live? 
    \\
    \textit{P21}: No, no. I lie a lot. I say I’m in an apartment, that I work well...It’s easier to make them believe I’m doing well than to show them I’m suffering. I’m not suffering that much, but...I don’t want them to see me like that, like I’m a failure.
\end{quote}
Some participants were more worried about having their families worry over them while so far away, which led P1 to not tell their family what was happening even whilst they were in the hospital:
\begin{quote}
    \textit{P1}: So far I haven't told anyone about my friends just three or four but not with too much detail no they don't even know that I'm here it's the friends that are on social media they don't understand that I'm in the United States. They don't know that I'm here, no, just a friend and an uncle and a grandmother.
\end{quote}
This silence illustrates how non-use can be protective on both ends: guarding individuals from questions and shielding families from fear, while also protecting one's privacy.

\subsubsection{Deletion Practices}
Deletion practices were widespread, especially when approaching borders. Participants described deleting contacts and messages to avoid scrutiny by smugglers, cartels, and law or immigration enforcement agents. Many explained that these practices stemmed from fears of being targeted for exploitation by desperate travelers, locals, cartel members, guides, and corrupt government officials. Others, particularly those who engaged in lending their mobile devices, deleted content to avoid being unfairly linked to potential criminal activity. 

Throughout the journey, many people on the move described memorizing loved ones' phone numbers and deleting contact information following phone use, especially when borrowing a device. P16 explained that even replying to a US-coded number (+1) could be dangerous, since it signaled potential access to money:
\begin{quote}
    \textit{P16}: Let’s say I get a message from my old boss or family, and I reply, forgetting to delete the number. Then someone checks the phone and sees the plus-one (country code), they might say, `This guy messaged someone in the US, we can make some money.' Then they call their people, like their friends, `Hey, this guy is walking around the center,' and now they’re following me. They’ll say, `I saw him already,' and next thing you know, I get kidnapped. They message my family and send pictures, `We have your son, your cousin, whoever, you have to pay. You need to send this much money so we let him go.' That’s how people get scammed. 
\end{quote}
Here, deletion was a form of anticipatory protection, preventing future exploitation. Others highlighted how deletion was a part of border crossing strategy. 
\begin{quote}
\textit{P13}: They told us that in Mexico they [immigration enforcement] were going to check our phones, for whom one communicated to, to see how they get to Mexico and all that. And then they were doing that with me. We deleted everything we had on the phone to be able to pass.
\end{quote}
For P13, erasure was not optional but necessary to move forward, showing how non-use was enforced by rumor, fear, and anticipated surveillance. 

\subsection{Coping Skills During Periods of Digital Disconnection} 
Since digital disconnection was viewed as inevitable, many participants developed coping strategies to endure and adapt. One common precaution was carrying inexpensive or older-model phones rather than newer, high-value models such as iPhones, which made them less desirable targets for theft (P14). Even so, periods without phones were unavoidable. In these moments, people on the move relied on communal practices and small acts of vigilance to maintain limited contact.

Borrowing phones was one of the primary strategies participants used to manage disconnection. Many explained that they would quickly reach out to those who are waiting for them in the US, before deleting the numbers they had used. This deletion reflected a constant awareness of risk: participants feared that the phone owner might exploit those numbers for ransom, or that they themselves could be endangered if the information fell into the wrong hands. Non-use here was a protective act, a way for participants to assert control over their safety even while improvising connectivity.

P32 recalled losing their phone mid-journey and enduring a month of disconnection. During that time, borrowed devices provided only minimal reassurance:
\begin{quote}
    \textit{P32}: Along the way I always met people from my country who were going to Peru, also people from Colombia who were going there, and I would borrow a call, they would give me a call or give me a message and the only thing I would tell them was‚ 'I’m fine'.
\end{quote}
For P32, and many others, the borrowed calls were ways of confirming health and survival across the precarious journeys. During these minimal communications, connection was reduced to its most essential form: letting others know that they were alive. Taken together, these coping practices frame non-use as resilience rather than deficit. 

%% file: sections/5_discussion.tex
\section{Discussion}
In this paper, we examine how people experiencing migration understand and experience unintentional ICT non-use (RQ1), how and why some intentionally avoid or refuse ICTs and sociotechnical systems (RQ2), and what practices, strategies, or workarounds people experiencing migration develop when ICT use is unavailable, inaccessible, or unsafe (RQ3).

Our data show that ICT engagement is not reducible to access or literacy; instead, it reflects patterned responses to changing legal, relational, and infrastructural conditions. To interpret these patterns, we draw on Ruthven’s information behavior theory of transitions, which explains major life disruptions as dynamic processes involving three distinct phases: \textit{understanding, negotiating, and resolving}~\cite{ruthven2022information}. Each of these phases is associated with different informational needs, vulnerabilities, and behaviors, and our participants described experiencing different forms of ICT non-use across these phases. We first situate the results within Ruthven's three phases of transitions, as well as Satchell and Dourish's typology of non-use---lagging adoption, disinterest, displacement, disenchantment, disenfranchisement, and active resistance~\cite{satchell2009beyond}. We then discuss reimagining sociotechnical systems in ways that expect digital disconnection.  

\subsection{Non-Use Across Transition Phases}
\subsubsection{Non-Use During the Understanding Phase}
Understanding is the sense-making process where individuals construct initial models of what migration may entail~\cite{ruthven2022information}. During this period, all of our participants described using ICTs to search for information by searching for or contacting people who have previously migrated, evaluating possible routes, or identifying perceived risks and opportunities. The few instances of ICT non-use in this phase of transition were related to \textit{disenfranchisement} through low comfort with technology (P2, P8, P9, P16, P30) or participants who experienced \textit{disenchantment} with systems due to encountering too much misinformation (P14, P21). A few participants described \textit{actively resisting} ICTs during this period in an effort to avoid information that might dissuade them from beginning the journey (P1, P4, P12). 


\subsubsection{Non-Use During the Negotiation Phase}
The negotiation phase of transitions is characterized by individuals managing their situations in ways that minimize disruption to their lives~\cite{ruthven2022information, jung2025transitions}. The vast majority of our participants, for example, had lived in numerous countries with higher linguistic,  cultural, and geographic proximity to their countries of birth prior to deciding to come to the US. The negotiation phase is enacted through travel, logistical coordination, creating connections with individuals of shared backgrounds and interaction with various legal and informal systems.

Most of the unintentional non-use our participants endured happened during the negotiation phase. In order to cope with their digital \textit{displacement} while traveling through remote areas, for example, every participant borrowed devices to contact loved ones or sponsors and receive information or resources along the way. In numerous instances, participants described experiencing \textit{disenchantment} during the negotiation phase, notably with the now-defunct CBP One process, where some participants reported that they had depleted resources waiting for a designated slot to enter the US, leading them to continue their journeys without securing appointments through the application, despite intending to utilize the system. Throughout our interviews, every participant expressed being aware of their vulnerability, and their susceptibility to \textit{disenfranchisement}, either through potential digital extortion by cartels and kidnappers (P5, P28), through exploitative fees for internet connection (P5, P10, P11, P12, P25, P30), or through federal entities withholding translation tools (P3, P4, P5, P17), information (P5, P21, P27), or ICTs entirely (all participants that encountered US DHS). Finally, \textit{active resistance} often looked like precautionary digital measures prior to crossing borders: deleting contacts and chats (P13, P14, P33), memorizing maps or law enforcement behavior before turning off digital devices (P10, P11, P16), or, in some cases, avoiding formal migration systems altogether (P15, P16, P20). 

Participants often anticipated periods of digital disconnection during their journeys, and this sometimes led to more intentional forms of ICT non-use during the negotiation phase, particularly \textit{lagging adoption} by not maintaining digital backups (P23, P24, P25), or forgoing the purchase of SIM cards while moving through small countries (most participants). Because they expected to be disconnected regardless, participants described choosing to save time and money rather than invest in connectivity. 

\subsubsection{Non-Use During the Resolving Phase}
In the resolving phase of transitions, individuals reconstruct and stabilize their orientations towards embodied interaction. Most participants navigated this phase through substantial \textit{use} of ICTs in order to comply with US institutional regimes. 

Most of the non-use during the resolving phase were unintentional, particularly through \textit{disenfranchisement}. Many participants shared stories of attempting to comply with US immigration systems but misunderstanding nuanced technical details about these systems, which led to missed requirements or penalties (P12, P13, P14, P21). 

\input{figures/designImplications2}
Intentional ICT non-use during the resolving phase often reflected efforts to preserve mental health and achieve stability. Some participants became \textit{disenchanted} with ICTs, withdrawing from social media and news consumption due to mental health concerns (P4, P18, P19, P24, P25), or withholding information from their families about their living conditions in the US (P1, P21). Similarly, a few participants expressed \textit{disinterest} with taking digital precautionary measures, saying they wanted to live as normal a life as possible (P13, P14). While rare, a few participants described \textit{active resistance} to online visibility through social media or an internet presence (P6, P17), or to visibility within government systems altogether (P20). 

Across these phases, non-use shifts according to what information is needed, what consequences are imagined, and what identities must be defended. This extends HCI work on non-use by showing that the absence from systems is a temporal phenomenon shaped by changing circumstances, risks, and institutional demands. 

\subsection{Reimagining Systems Beyond Infrastructures that Depend on Non-Use}
Our findings suggest that ICT non-use is a structural feature of contemporary migration infrastructures. Systems governing mobility, asylum, and resettlement depend on continuous digital presence for identification, coordination, and compliance, yet systematically place people in conditions where such presence cannot be maintained. Non-use emerges from this contradiction: visibility is required, while the material means of remaining visible are unstable, inaccessible or actively removed. 

Participants' accounts illustrate the material vulnerability of digital technologies~\cite{latonero2018digital}. Similar to Fisher et al.'s study of the Za'atari refugee camp in Jordan~\cite{Fisher2017Za'atari}, those crossing the US/MX border have limited access to technology and what little access they have is often inoperable or costly. Devices are lost during travel, confiscated at checkpoints, damaged by environmental exposure, or rendered unusable through cost and geographic isolation. Phones become central points of failure within systems that treat them as durable anchors of identity and access. This produces chronic disenfranchisement~\cite{satchell2009beyond} through what Larkin calls infrastructural precarity, the foreclosure of possibility that occurs when people can no longer act within the systems that organize their lives~\cite{larkin2013politics}. Periods of disconnection, therefore, reorganize what forms of participation, recognition, and mobility remain possible. 

The effects of non-use extend across time and institutions. Participants described secondary consequences following device loss or lack of connectivity, including missed legal appointments, loss of identity records, inability to locate family members, and unintentional entry into legal violation. When digital presence collapses, legibility to state systems collapses with it. Non-use becomes a mechanism through which people are sorted out of bureaucratic timelines and excluded from processes that require continuous traceability.

At the same time, participants actively work within these conditions through intentional non-use. Avoiding platforms, deleting data, and limiting online visibility function as protective practices against surveillance, extortion, and detention. These strategies are tactical responses to infrastructural risk. In such cases, participants balanced their visibility with perceived safety. 

Reimagining systems beyond infrastructures that depend on non-use requires designing for this reality rather than against it. Following Lloyd et al.~\cite{lloyd2013connecting}, responsibility of connectivity must shift from individuals to service providers. Connectivity should be treated as infrastructure rather than as a commodity, particularly in camps, shelters, border crossings, and detention centers. Systems must remain usable under conditions of loss, refusal, and disconnection. 

This implies design approaches that assume intermittent presence as a baseline condition. Offline-first interaction, communal devices, resumable workflows after device loss or detention, and participation without persistent identification become central design requirements rather than edge cases. Disconnection should be \textit{seamful}; treated as an expected state that systems are structured around rather than as an exception to be patched over~\cite{inproceedings}. 

Across accounts, non-use reveals how power operates through infrastructure. Participants recognize that effort, intention, and even compliance were often insufficient to overcome structural constraints. Access to safety, documentation, and mobility depends on systems that require ongoing digital visibility, even as those same systems made it difficult to maintain. Designing beyond infrastructure that depends on non-use, therefore, means creating systems that continue to function when people become disconnected, and that support agency, identity, and access even during periods of absence.  


\subsection{Implications: Designing \textit{for} Disconnection}
This work suggests a shift in how sociotechnical systems are designed for contexts of migration: disconnection must be treated as an ordinary operating condition. Across participants' experiences, periods of absence caused by issues like device loss, confiscation, detention, network instability, or intentional withdrawal, were recurring and predictable. Design implications, therefore lie less in refining interaction and more in establishing infrastructural commitments that allow systems to remain usable despite interruption.

Table~\ref{tab:design_implications2} articulates design commitments derived from observed practices of ICT non-use across migration transitions. The table frames non-use as a predictable and often rational response to legal, infrastructural, and relational conditions. The commitments articulated here reorient sociotechnical design away from assumptions of continuous presence, visibility, and optimization, and toward persistence, recoverability, and access that remains viable across absence, withdrawal, and refusal.

Designing for disconnection also raises questions about how mobility, governance, and borders are encoded into sociotechnical infrastructures. Capital circulates across borders with few interruptions, while people in migration remain bounded by systems that require national authentication and uninterrupted visibility. In this context, infrastructures designed around nation-state verification reproduce many of the conditions that generate unintentional non-use. Systems that can function when visibility to the state is absent (through distributed verification, mediated access, portable identity, etc.) better align with how people actually move through transnational spaces. Importantly, these approaches do not eliminate governance, but shift responsibility away from those who are already structurally disadvantaged.

Designing for non-use is not designing for failure; it is designing for the world as it is lived. Systems that remain functional during disconnection can support people's agency, identity, rights, and access, even when digital presence is unstable or temporarily impossible.

%% file: figures/designImplications2.tex
\begin{table*}[t]
\centering
\caption{Designing for Disconnection: Principles Derived from Practices of ICT Non-Use}
\label{tab:design_implications2}
\begin{tabular}{p{2.25cm} p{3.4cm} p{4cm} p{3cm} p{3.4cm}}
\toprule
\textbf{Type(s) of \newline Non-Use} & \textbf{Design Principle} & \textbf{Non-Use Practice \newline Observed}  & \textbf{Design Assumption Addressed} & \textbf{Example} \\

\midrule

\midrule
Displacement &
Treat absence as predictable  &
Device loss, confiscation, detention, or reliance on borrowed devices interrupts access &
Users maintain stable proximity to personal devices &
Offline-first workflows that save progress and sync when connectivity returns\\

\midrule
Disenchantment &
Preserve eligibility across absence; Absence should incur inquiry rather than penalty &
Participants disengage due to misinformation, prolonged delay, or emotional exhaustion &
People have unlimited resources; more information always improves outcomes &
Wait-time sensitive systems; Pausable interactions or participation without loss of status \\

\midrule
Disenfranchisement &
Connectivity and legibility are institutional responsibilities &
Loss of access due to cost, language barriers, withheld tools, or infrastructural failure &
Access and literacy are individual responsibilities &
Shared access points as institutional infrastructure (e.g., translation tools) \\

\midrule
Lagging adoption &
Access must not depend on early or continuous enrollment; Design for delayed or partial adoption &
Participants forgo SIM cards, backups, or app installation because disconnection is expected &
Adoption reflects readiness or commitment &
Services accessible without prior registration or persistent accounts;  proxy access via NGOs or trusted entity \\

\midrule
Disinterest &
Systems should function without requiring continual self-management &
Participants reject ongoing precautionary digital practices to live “normally” &
Responsible users constantly optimize safety and preparedness &
Services usable without repeated security checks or configuration \\

\midrule
Active resistance &
Refusal and selective visibility must remain compatible with access &
Participants delete data, limit traceability, or avoid platforms to reduce risk &
Visibility and persistence are inherently beneficial &
Selective identity credentials not tied to
a single authority \\

\bottomrule
\end{tabular}
\end{table*}

%% file: sections/6_conclude.tex
\section{Conclusion}
This paper has shown that people on the move experience ICTs in ways that cannot be understood solely through adoption and use. Non-use---both enforced absence through device loss, detention, or infrastructural denial, and deliberate refusal as a survival tactic---shapes how people navigate surveillance, uncertainty, and systemic precarity. Across Ruthven’s three transition phases, we observed dynamic, patterned practices of non-use, which when interpreted through Satchell and Dourish's typology, reveal that non-use is not a failure to connect but a situated response to power, precarity, and control. For HCI, these findings expand calls to design for disconnection, which should be expected orientations within sociotechnical systems rather than as punitive anomolies. 

More broadly, this work underscores the political stakes of HCI research. Encounters with ICTs reveal how infrastructures encode assumptions about stability, compliance, and trust---assumptions that collapse during the migration journey. Designing for non-use requires reimagining technologies that do not punish absence, that recognize refusal as legitimate, and support dignity, agency, and survival. CHI researchers and designers must therefore build systems that are not only technically robust but also ethically accountable, attending to how infrastructures can either reproduce oppression or create space for resilience and human flourishing.

\begin{acks}
We are profoundly grateful to the participants---many now deported, repatriated, detained, or otherwise facing ongoing uncertainty---for their courage, transparency, trust in us, and for making this research possible. We are additionally grateful for the help and feedback from our community partners and friends across the aid network of El Paso, TX, USA and Cd. Juárez, CH, MX. This work would not have been possible without the dedication, commitment, and care of our interview team: Gabriela Ortiz, Aarón Argüello Labandera, and Emilia Rosas. This manuscript has improved significantly thanks to the helpful feedback of anonymous reviewers, Jessica Needle, as well as Bill Hamilton and the PLEX lab. 
\end{acks}

%% file: sections/7_appendix.tex
\newpage
\section{Appendix A: Semi-Structured Interview Protocol Question Themes:}
\label{appendix:protocol}
\begin{itemize}
        \item Introductions, ICT use
        \begin{itemize}
            \item Information ecosystems
            \item Social media usage
            \item News and information sources
            \item Communication with others
        \end{itemize}

        \item Information Preparation
        \begin{itemize}
            \item If and where (online or in person) participants sought information
            \item How they sought help (people-based, digital, or mixed) for ICT challenges
            \item How they maintained a sense of community
            \item How they maintained communication with loved ones
            \item Types of information shared
            \item How new connections were made
            \item Tech-based precautions to protect privacy
        \end{itemize}

        \item Border Encounter and Processing Experiences
        \begin{itemize}
            \item Entry through different countries
            \item Encounters with powerful stakeholders (migration, military, police, NGOs, NPP officials)
            \item What kind of information was collected
            \item Perceived purpose of collected information
            \item Experiences with device loss and digital disconnection
            \begin{itemize}
                \item Confiscation or return of belongings
            \end{itemize}
            \item Clarity of instructions from immigration officers
            \item Language access while entering a country
            \item Understanding outcomes across administrative stages
            \item Monitoring devices or migration systems used (e.g., CBP One)
            \begin{itemize}
                \item Functions of the tech
                \item Type of info collected
                \item Role in daily routines
                \item Disruptions to their lives
                \item Perceptions of purpose and impacts
            \end{itemize}
        \end{itemize}

        \item Comfort with Technology and Information Preparedness
        \begin{itemize}
            \item Access to translation or interpretation resources
            \item Exposure to misinformation or fraud
            \begin{itemize}
                \item Comparison of online migration narratives and lived experiences
            \end{itemize}
        \end{itemize}

        \item Dataveillance
        \begin{itemize}
            \item Device and app usage (phone type, apps, accounts)
            \item What they believe each platform knows about them
            \item What they watch on each platform
            \item Public/private nature of accounts
            \item Nature of use or non-use
            \item What information participants believe their phone contains
            \item Purpose of information work
            \begin{itemize}
                \item If struggling: ask how Maps works
                \item Reconnect to social media data afterward
            \end{itemize}
        \end{itemize}

        \item Concluding thoughts
        \begin{itemize}
            \item Sentiments about the future
            \item Messages to powerful stakeholders
            \item Do you feel “watched”?
        \end{itemize}
    \end{itemize}